\documentclass[preprint,aps,10pt]{revtex4}
\usepackage{amsmath,amssymb,amsthm,graphicx,latexsym}
\newcommand{\doublespacing}{\let\CS=\@currsize\renewcommand{\baselinesstrech}
{2.0}\tiny\CS}

\newcommand{\bd}{\begin{document}}
\newcommand{\ed}{\end{document}}
\newcommand{\bc}{\begin{center}}
\newcommand{\ec}{\end{center}}
\newcommand{\vs}{\vspace}

\begin{document}

\title{\large \bf Spectral Singularity in confined ${\cal{PT}}$ symmetric optical potential}

\vs{2cm}

\author{{\bf Anjana Sinha}$^*$ and {\bf R. Roychoudhury}$^ {\#}$  }

\vspace{2cm}

\begin{abstract}

\vspace{.5cm}

\noindent We present an analytical study for the scattering amplitudes (Reflection $|R|$ and Transmission $|T|$), of the periodic ${\cal{PT}}$ symmetric optical potential $ V(x) = \displaystyle  W_0 \left( \cos ^2 x +  i V_0 \sin 2x \right) $
confined within the region $0 \leq x \leq L$, embedded in a homogeneous medium having uniform potential $W_0$. The confining length $L$ is considered to be some integral multiple of the period $ \pi $. We give some new and interesting results. Scattering is observed to be normal ($|T| ^2 \leq 1, \ |R|^2 \leq 1$) for $V_0 \leq 0.5 $, when the above potential can be mapped to a Hermitian potential by a similarity transformation. Beyond this point ($ V_0 > 0.5 $) scattering is found to be anomalous ($|T| ^2, \ |R|^2 $ not necessarily $ \leq 1 $). Additionally, in this parameter regime of $V_0$, one observes infinite number of spectral singularities $E_{SS}$ at different values of $V_0$. Furthermore, for $L= 2 n \pi$, the transition point $V_0 = 0.5$ shows unidirectional invisibility with zero reflection when the beam is incident from the absorptive side ($Im [V(x)] < 0$) but finite reflection when the beam is incident from the emissive side ($Im [V(x)] > 0$), transmission being identically unity in both cases. Finally, the scattering coefficients $|R|^2$ and $|T|^2 $ always obey the generalized unitarity relation : $ ||T|^2 - 1| = \sqrt{|R_R|^2 |R_L|^2}$, where subscripts $R$ and $L$ stand for right and left incidence respectively.

\vs{.5cm}


\vs{.5cm}


\vspace{1cm}

\noindent $^*$ Department of Instrumentation Science, Jadavpur University, Kolkata - 700 032, INDIA \\
e-mail : anjana23@rediffmail.com; sinha.anjana@gmail.com

\vs{.3cm}

\noindent $^{\#}$ Visiting Professor, Dept. of Mathematics, Bethune College, Kolkata - 700 006, INDIA \\
\& Advanced Centre for Nonlinear and Complex Phenomena, 1175 Survey Park,  Kolkata - 700075, INDIA \\
e-mail : rajdaju@rediffmail.com; rroychoudhury123@gmail.com \\

\vs{.3cm}

\noindent fax : +91 33 24146321; phone : +91 33 25753020

\end{abstract}

\maketitle

\newpage

\section{Introduction}

Ever since the experimental verification of ${\cal{PT}}$ symmetry in optical structures \cite{pt-expt1,pt-expt2,pt-expt3},
complex optical potentials have been the subject of much attention for the past 5-6 years or so \cite{pt-opt1,pt-opt2,pt-opt3,pt-opt4,pt-opt5,pt-opt6,pt-opt7,pt-opt8,longhi-ss,mostafazadeh-ss,jones-jpa,plyushchay}.
This is primarily because of the mathematical isomorphism
between the quantum mechanical Schr\"{o}dinger equation and the
paraxial equation of propagation of an electromagnetic wave in a medium, viz.,
\begin{equation}\label{opt}
    \displaystyle i \frac{\partial \psi}{\partial z} = - \left\{ \frac{\partial ^2 }{\partial x^2} + V(x) \right\} \psi =
    k^2 [ 1 + 2 v(x)] \psi = E \psi
\end{equation}
where $\psi (x, z)$ represents the envelope function of the amplitude of the electric field, $z$ is a
scaled propagation distance which plays the role of time $t$ in quantum mechanics, $x$ denotes the spatial coordinate, and $V (x) = V(x+D)$ is the optical potential of period $D$, proportional to the refractive index $n(x)$ of the material through which the wave is passing. A complex $V(x)$ corresponds to a complex refractive index $n(x) = n_0 + n_R(x) + n_I (x)$, where $n_0$ represents the constant substrate background index, with $n_I$ being associated with gain or loss. Additionally, $n_{R,I} (x) \ll n_0$ . To be more precise, $ V(x) = n_0 [1 + v(x)]$, with $|v| < < 1$. For ${\cal{PT}}$ symmetry, the gain and loss regions need to be carefully configured so that $V(x) = V^* (-x)$, {\it i.e.} the real part $n_R (x) $ should be even and the imaginary part $n_I (x)$ odd. Eq. (\ref{opt}) describes Bragg scattering of matter waves from a complex potential in the non-interacting regime, e.g., dilute cold atomic beam. It may be mentioned that the complex potential arises from the interaction of near resonant light with an open 2-level system. Thus Bragg scattering of optical waves in 1-d Bragg grating structure is similar to scattering of matter waves in the framework of eq. (\ref{opt}).

Two of the most prominent features explored in Non Hermitian quantum systems are : \\
~ (i) {\it Exceptional points} (EP) in the discrete spectrum, also referred to as
Hermitian degeneracies, where both eigenvalues and eigenvectors coalesce with the variation of a system parameter. \\
(ii) {\it Spectral singularities} (SS) in the continuous part of the spectrum, which
correspond to resonance states with vanishing spectral width.

{\it Exceptional points} and {\it Spectral singularities} are mathematical concepts with intriguing physical realizations.
While energy values switch from real to complex conjugate pairs at an EP, reflection and transmission coefficients blow up at SS. In various studies on complex crystals with periodic potentials, the physical implications of spectral singularities have been investigated \cite{longhi-ss,mostafazadeh-ss,jones-jpa,zafar-2013}. SS, which are actually lasing thresholds, are known to spoil the completeness of Bloch-Floquet states in non-Hermitian Hamiltonians with complex periodic potentials. The onset of spectral singularities in complex crystals (occurring at the ${\cal{PT}}$ symmetry breaking point) is found to be associated with secularities that arise in Bragg diffraction patterns. Experimentally, spectral singularities can be revealed in diffraction experiments on non Hermitian systems with complex potentials, by the appearance of a secular growth of the
amplitudes of waves scattered off the lattice when it is excited
by a plane wave at special incident angles. Recently Mostafazadeh has shown that nonlinear spectral singularities are intensity dependent, corresponding to emission of waves with a particular wavelength-amplitude profile \cite{mostafa-prl-2013, mostafa-pra-2013}. In particular, for a Kerr nonlinearity, the author showed that the first order perturbative equation that determines the nonlinear SS provides an explicit expression for the intensity of emitted waves from an infinite planar slab of gain medium.

In most of the studies done on optical potentials so far \cite{pt-opt1,pt-opt2,pt-opt4,longhi-ss,jones-jpa,bikash-PLA,ge-pra}, the general interest lies in obtaining the band structure for complex periodic potential of the type
\begin{equation}\label{schro-opt}
    \displaystyle V(x)  = \displaystyle W_0 \left\{ \cos ^2 x + i V_0 \sin 2x \right\}
\end{equation}
with period $D = \pi $, where $W_0$ represents the lattice amplitude and $V_0$ is a measure of the strength of the non Hermitian part of the potential, i.e., the gain-loss periodic distribution. For the Hermitian case, there is no gain-loss modulation, and $V_0 = 0$. The reason for the widespread interest in the particular potential given in eq. (\ref{schro-opt}) above, lies in the fact that it gives many physically interesting results, especially highlighting unusual
diffraction and transport properties of complex optical lattices. In various studies for this potential, using spectral techniques and detailed
numerical calculations, the ${\cal{PT}}$ threshold $V_0 ^{th} = 0.5$ has been identified,
below which all eigenvalues for every band and every Bloch wave number are real (unbroken ${\cal{PT}}$ phase)
and all the forbidden gaps are open, whereas at the threshold $V_0 ^{th} = 0.5$, the band gaps
vanish \cite{pt-opt1,pt-opt2,longhi-ss}. On the other hand, beyond this value (i.e., $V_0 > 0.5$) ${\cal{PT}}$ symmetry is spontaneously broken, energies turn complex and the bands start merging together forming oval-like structures.
Subsequently, in an analytic study on this particular potential \cite{bikash-PLA}, the authors found the existence of a second critical point at $V_0 = 0.888437$, beyond which no part of the band structure remains real.

Our aim in this work is to look for the effect of the ${\cal{PT}}$ phase transition on the continuous part of the spectrum. Motivated by the physical importance of SS mentioned above, we shall focus our attention on the scattering amplitudes --- reflection $|R|$ and transmission $|T|$ --- in our effort to search for spectral singularity. For this purpose, we shall consider an optical structure having a ${\cal{PT}}$ symmetric potential given in eq. (\ref{schro-opt}) above, confined between two bounding walls at $x=0$ and $x= L$, embedded in a homogeneous medium having uniform potential $W_0$; i.e., $V(x) = W_0$ for $x \leq 0$ and $x \geq L$. It may be mentioned that in \cite{longhi-ss} the author carried out an asymptotic analysis of the above potential in the entire crystal for the particular value $V_0 = 0.5$, and showed the existence of an infinite number of spectral singularities in terms of the secular growth of the wave amplitude. In \cite{jones-jpa} analytical expressions were obtained for the reflection and transmission coefficients for $V_0 = 0.5$, in terms of modified Bessel functions. While examining unidirectional invisibility ---  identically unit transmission, zero reflection from one side and enhanced reflection from the other --- they investigated how the enhanced reflection from one side comes about from a wave packet as opposed to a plane wave.  In yet another study \cite{pt-opt4}, the authors considered bounding walls at $x=\pm L/2$, and showed the phenomenon of unidirectional reflectivity. However, our present work significantly differs from these earlier studies : while ref. \cite{longhi-ss} and \cite{jones-jpa} consider the potential to be valid in the entire region, we consider bounding walls at $x=0$ and $x=L$. Additionally, these studies are for the critical value $V_0=0.5$ only. We consider all possible values of $V_0$, both above, below and at the critical point. Regarding ref. \cite{pt-opt4}, though the potential considered there is similar to that studied in this work, nevertheless, ours is a direct method as against their numerical simulation; we determine analytical expressions for the reflection and transmission coefficients, in terms of Mathieu functions and their derivatives.

\vs{.2cm}

The article is organized as follows : In Section 2, we briefly discuss the different parameter regimes of $V_0$, to obtain the solutions in the different cases. Applying the boundary conditions, we determine the reflection and transmission coefficients $|R|^2$ and $|T|^2$ respectively, for $L = n \pi$, where $n$ may be even or odd integer, including the case of a single period $n=1$, for waves incident from the left as well as right. Our observations are plotted in Figures 1 to 7 --- the real and imaginary parts of $V(x)$ are plotted in Fig. 1, whereas $|R_{R,L}|^2$ and $|T|^2$ for different $L$ are plotted in Fig. 2 ($V_0 < 0.5$), Fig. 3 ($V_0 > 0.5$) and Fig. 6 ($V_0 = 0.5$). The 3-d plot of Fig. 4 depicts the blowing up of $|T|^2$ at innumerable spectral singularities, for different sizes of the optical structure $L$. Fig. 5 gives some typical values of energy $E_{SS}$ where SS occurs for different $V_0$ and $L$. Fig. 7 is for the Hermitian case when $V_0 = 0$. Finally, our observations are summarized in Section 3.

\section{Theory}

We start with the linear equation
\begin{equation}\label{schro-eq}
    \displaystyle \left\{ \frac{\partial ^2 }{\partial x^2} + V(x) \right\} \psi = \beta \psi
\end{equation}
where $\beta$ represents the propagation constant in the periodic
structure, and
\begin{equation}\label{pot}
        V(x) = \left\{
    \begin{array}{lcl}
        & & \displaystyle W_0 \left( \cos ^2 x + i V_0 \sin 2x \right)
        , \  0 < x < L   \\ \\
        & & \displaystyle W_0 , \qquad \  - \infty <  x < 0 , \ L < x < \infty  \\
    \end{array}
    \right.
\end{equation}
In the different parameter regions of $V_0$, the potential in eq. (\ref{pot}) may be written as
\begin{equation}\label{v}
    V(x) = \left\{
    \begin{array}{lcl}
        & & \displaystyle \frac{W_0}{2} \left\{ 1 + U_1 (x) \right\}  , \   V_0 < 0.5    \\ \\
        & & \displaystyle \frac{W_0}{2} \left( 1 + e^{2 i x} \right)  , \ \ \ \  V_0 = 0.5  \\ \\
        & & \displaystyle \frac{W_0}{2} \left\{ 1 + i U_2 (x) \right\}   , \  V_0 > 0.5    \\ \\
    \end{array}
    \right.
\end{equation}
where
\begin{equation}\label{u1x}
    U_1 (x) = \displaystyle \sqrt{1 - 4 V_0 ^2} \ \cos \left\{
        2 x - i \tanh ^{-1} \left( 2 V_0 \right) \right\}
\end{equation}
\begin{equation}\label{u2x}
    U_2 (x) = \displaystyle \sqrt{4 V_0 ^2 - 1} \ \sin \left\{
        2 x - i \tanh  ^{-1} \displaystyle \left( \frac{1}{2 V_0} \right) \right\}
\end{equation}
For a clear picture of the model considered here, we plot the real and imaginary parts of the potential in Fig. 1, for 2 different lengths --- $L= \pi$ and $L= 5 \pi$, for $W_0 = 4$, and three different values of $V_0$.

\vspace{.3cm}

\begin{figure}[ht]
\begin{center}
\includegraphics[width=4.5 cm,height=3.4 cm]{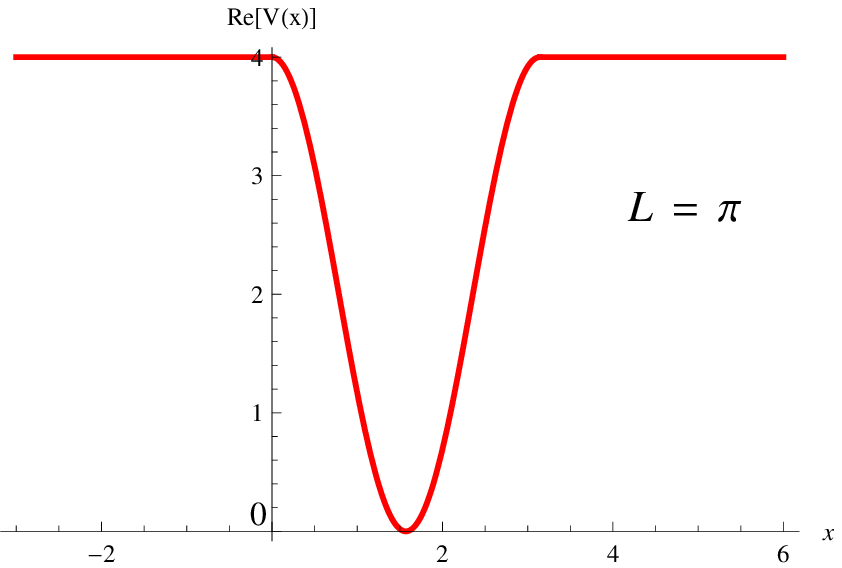} ~~~~~~~~~~~~~~~~~ \includegraphics[width=4.5 cm,height=3.4 cm]{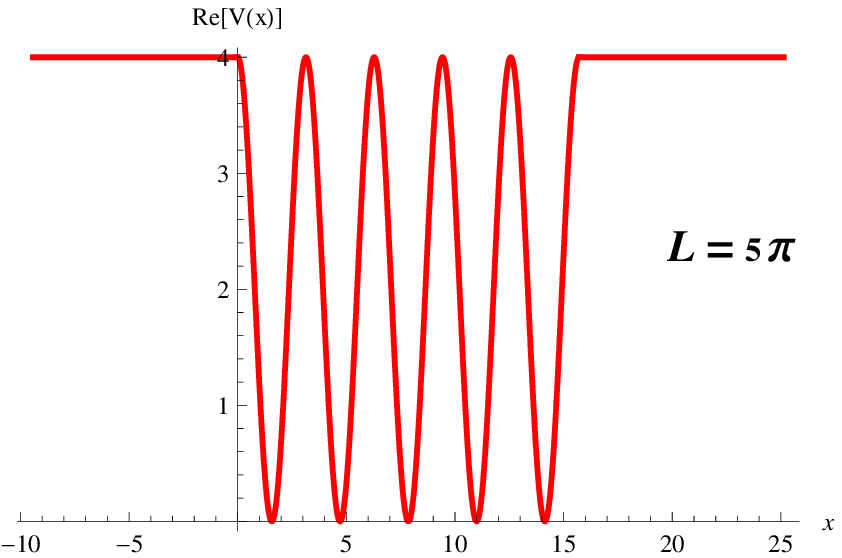} \\
\includegraphics[width=4.5 cm,height=3.4 cm]{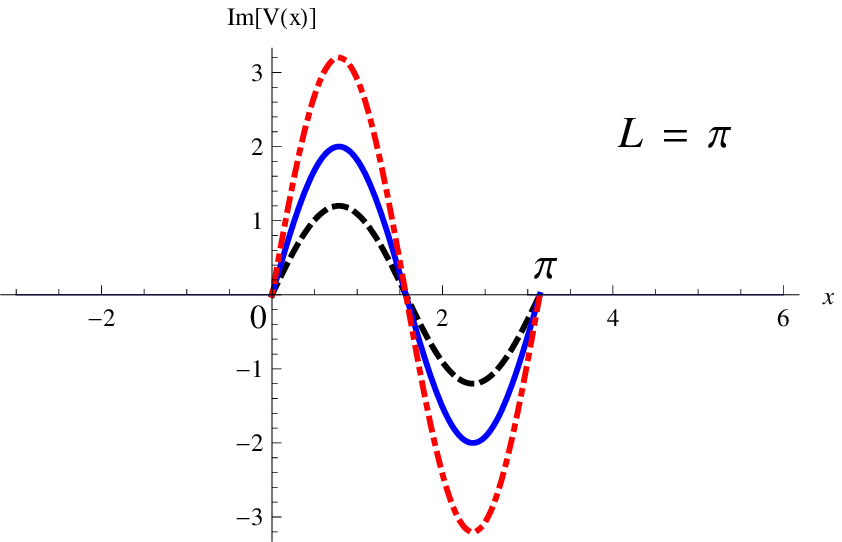} ~~~~~~~~~~~~~~~~~ \includegraphics[width=4.5 cm,height=3.4 cm]{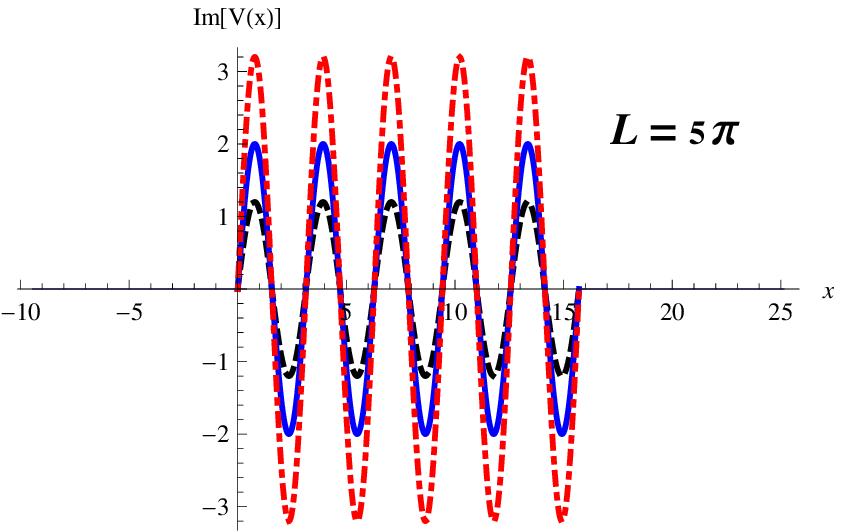} \\
\caption{{Color online : } ~ (a) {\small Plot showing $Re[V(x)]$ in unit ${\cal{PT}}$ cell $L = \pi$ , for $W_0=4$ } ~  (b)
    {\small Plot showing $Re[V(x)]$ for $L = 5 \pi$ and $W_0=4$ }  ~ (c) {\small Plot showing $Im[V(x)]$ for $L = \pi$ and  $W_0=4$ } ~  (d)
    {\small Plot showing $Im[V(x)]$ for $L = 5 \pi$ and $W_0=4$, ~~~~~  for $V_0 = 0.3$ (black dashed line), $V_0 = 0.5$ (blue solid line),
    $V_0 = 0.8$ (red dotdashed line)}  }
\end{center}
\end{figure}

\vspace{.3cm}

Outside the bounding walls at $ x = 0 $ and $ x = L $, the
solutions in the two regions, for wave incident from left, are
\begin{equation}\label{psi-out}
\begin{array}{lcl}
    \psi _L (x)  &=& \displaystyle e^{i k x} + R e^{-ik x} \ , \ \ x < 0 \\ \\
    \psi _R (x)  &=& T e^{i k x} \ , \ \qquad \qquad x > L \\
\end{array}
\end{equation}
where $R$ and $T$ denote the reflection and transmission
amplitudes, and
\begin{equation}\label{k}
    k = \displaystyle  \sqrt{ \left( E  - W_0 \right) }
\end{equation}
in units $\hbar = c = 2m = 1$. (The energy $E$ is related to $\beta$.)
The next step would be to find the solutions within the ${\cal{PT}}$ cell for different parameter values of $V_0$.

\vspace{1cm}

\noindent {\bf Case 1 : $V_0 < 0.5$}

\vs{.2cm}

If one changes the variable to
\begin{equation}\label{y-less}
    y = x - \displaystyle \frac{i}{2} \ \tanh  ^{-1} (2 V_0)
\end{equation}
then eq. (\ref{schro-eq}) reduces to the Mathieu equation
\begin{equation}\label{math-cos}
    \displaystyle \frac{d^2 \psi}{dy^2} + \left[ a - 2q \ \cos 2y \right] \psi = 0
\end{equation}
with characteristic value
\begin{equation}\label{a-cos}
    \displaystyle a =  \frac{W_0}{2} - \beta
\end{equation}
and
\begin{equation}\label{q-cos}
    \displaystyle q =  \frac{W_0}{4} \sqrt{1 - 4 V_0 ^2}
\end{equation}
Thus, within the complex periodic structure, the solutions to eq. (\ref{math-cos}) are given in terms of the Mathieu functions
\begin{equation}\label{psi-cos}
    \displaystyle \psi _{in} (y) = A_1 \ Me _{\nu} ^1 \left( a,q,y \right) \ + \ A_2 \ Me _{\nu} ^2 \left( a,q,y \right)
\end{equation}
where $ Me _{\nu} ^1 \left( a,q,y \right) $ and $ Me _{\nu} ^2 \left( a,q,y \right)$ are the two independent solutions of the Mathieu equation, and $\nu$ is the characteristic Mathieu exponent. In the notation of Meixner and Sch\"{a}fke \cite{meixner},  \begin{equation}\label{me-nu}
    Me _{\nu} (y,q) = \displaystyle e^{i \nu y} \sum _{r = -
    \infty} ^{\infty}  c _{\nu , 2r} e^{2 r i y}
\end{equation}
The coefficients $c _{\nu , 2r}$ satisfy
\begin{equation}\label{c-nu}
  \begin{array}{rrr}
   \displaystyle q \ c _{\nu , 2r-2} + \left\{ (2r + \nu )^2 - a
    \right\}  c _{\nu , 2r} + q \  c _{\nu , 2r + 2} &=& 0 \\ \\
    \displaystyle \lim _{r \rightarrow \pm \infty} c_{\nu , 2r} &=& 0 \\
    \end{array}
\end{equation}
subject to the normalizing condition
\begin{equation}\label{c-normalize}
    \displaystyle \sum _{r = -
    \infty} ^{\infty}  \left( c _{\nu , 2r} \right) ^2 = 1
\end{equation}
It is to be noted that the limit in eq. (\ref{c-nu}) does not
automatically hold for arbitrary values of the parameters $q, \ a
$ and $\nu$, but rather only for a specific value or values of one
parameter when the other two are prescribed. Since we are
considering $L= m \pi$, we should consider the analytic continuation
of Mathieu functions, derived from its pseudo-periodicity property \cite{handbook}
\begin{equation}\label{mathieu-cont}
    F _{\nu} (\pm y + m \pi) = \displaystyle e^{\pm i m \nu \pi} F _{\nu} (\pm y)
\end{equation}
for fixed $a$ and $q$ (Floquet solution's property), with
    \begin{equation}
    F_{\nu} (y) = Me _{\nu} (a,q,y)
    \end{equation}
Now, if $\nu$ is not an integer, then $F_{\nu} (y)$ and $F_{\nu} (-y)$ are linearly independent, and the general solution of (\ref{math-cos}) can be written as
\begin{equation}\label{floquet}
    \psi = \displaystyle A_1 \ F_{\nu} (y) + A_2 \ F_{\nu} (-y)
\end{equation}
If $ A_1 A_2 \neq 0$, $\psi$ is not a solution of Floquet's equation. \\
If $\nu$ is an integer, then $F_{\nu} (y)$ and $F_{\nu} (-y)$ are linearly dependent. In this case the second independent solution is given by
\begin{equation}\label{floq-1}
   \psi _2 = \displaystyle y \ ce_{\nu} (y,q)  + \sum _{k=0} ^{\infty}  d_{2k +p} \sin (2k +p) y
\end{equation}
or
\begin{equation}\label{floq-2}
   \psi _2 = \displaystyle y \ se_{\nu} (y,q)  + \sum _{k=0} ^{\infty}  f_{2k +p} \cos (2k +p) y
\end{equation}
where $ce_{\nu} (y,q)$ and $se_{\nu} (y,q)$ are given by
\begin{equation}\label{cer}
    ce_{\nu} (y,q) = \displaystyle \frac{1}{2} \left[ F_{\nu} (y) + F_{\nu} (-y) \right]
\end{equation}
\begin{equation}\label{ser}
    se_{\nu} (y,q) = \displaystyle -i \ \frac{1}{2} \left[ F_{\nu} (y) - F_{\nu} (-y) \right]
\end{equation}
Thus $ce_{\nu} (y,q)$ and $se_{\nu} (y,q)$ are even and odd functions of $y$, respectively, for all $\nu$. If $\nu$ is an integer, then $ce _{\nu}$ and $se _{\nu}$ are either Floquet solutions or identically zero, so that the analytic continuation given in (\ref{mathieu-cont}) can be carried out; for details see \cite{handbook}.

\vspace{1cm}

In order to calculate the scattering amplitudes $R$ and $T$, applying the boundary conditions, we make use of the properties of Mathieu functions and their derivatives \cite{handbook}. This also gives the coefficients $ A_1, \ A_2$. For brevity, we just quote the results here
\begin{equation}\label{t}
    T = \displaystyle \frac{1}{2 i k} \ e^{-i k L}
    A_1 \left( i k M^1 (y_L) + M^{1 \ \prime} (y_L) \right) + A_2 \left( i k M^2 (y_L) + M^{2 \ \prime} (y_L) \right)
\end{equation}
\begin{equation}\label{rL}
    R_L = \displaystyle \frac{1}{2 i k } \  \left[ A_1 \left( i k M^1 (y_0) - M^{1 \ \prime} (y_0) \right) + A_2 \left( i k M^2 (y_0) - M^2 (y_0)  ^{\prime} \right) \right]
\end{equation}
\begin{equation}\label{rR}
    R_R = \displaystyle \frac{1}{2 i k \ e^{i k L}} \ \left[ A_3 \left( i k M^1 (y_L) + M^{1 \ \prime} (y_L) \right) + A_4 \left( i k M^2 (y_L) + M^{2 \ \prime} (y_L) \right) \right]
  \end{equation}
\begin{equation}\label{a1a2}
    A_1 = \displaystyle 2 i k \ \sigma _1   \ \ , \ \ A_2 = \displaystyle - A_1 \ \frac{i k M^1 (y_L) - M^{1 \ \prime} (y_L)}{i k M^2 (y_L) - M^{2 \ \prime} (y_L)}
\end{equation}
\begin{equation}\label{a3a4}
    A_3 = \displaystyle 2 i k e^{-ikL} \ \sigma _2 \ , \ A_4 = \displaystyle - A_3 \ \frac{ik M^1 (y_0) + M^{1 \ \prime} (y_0)}{ik M^2 (y_0) + M^{2 \ \prime} (y_0)}
\end{equation}
\begin{equation}\label{sigma1}
    \sigma _1 = \displaystyle \frac{ i k M^2 (y_L) - M^{2 \ \prime} (y_L) }{\left(i k M^2 (y_L) - M^{2 \ \prime} (y_L) \right) \left(i k C_0 + M^{1 \ \prime} (y_0) \right) -
    \left(i k M^1 (y_L) - M^{1 \ \prime} (y_L) \right) \left(i k M^2 (y_0) + M^{2 \ \prime} (y_0) \right)}
\end{equation}
\begin{equation}\label{sigma2}
    \sigma _2 = \displaystyle \frac{ ikM^2 (y_0) + M^{2 \ \prime} (y_0)}{\left( ikM^1 (y_L) - M^{1 \ \prime} (y_L) \right)
    \left( ikM^2 (y_0) + M^{2 \ \prime} (y_0) \right) - \left(ikM^2 (y_L) - M^{2 \ \prime} (y_L) \right) \left(ik M^1 (y_0) + M^{1 \ \prime} (y_0) \right)}
\end{equation}
where $M^1 (y_0) $ and $ M^1 (y_L) $ denote the values of Mathieu function $ Me _{\nu} ^1 \left( a,q,y \right) $ at $x=0$ and $x=L$, and $M^2 (y_0) $ and $ M^2 (y_L) $ represent the values of Mathieu function $ Me _{\nu} ^2 \left( a,q,y \right) $ at $x=0$ and $x=L$. $M^{1 \ \prime} (y_0),\ M^{1 \ \prime} (y_L), \ M^{2 \ \prime} (y_0) , \ M^{2 \ \prime} (y_L) $ represent the corresponding derivatives wrt $x$, evaluated at $x=0$ and $x=L$.

In Fig. 2, we plot the scattering coefficients as well as a typical scattering state, for $V_0 < 0.5$. In particular, the scattering coefficients are plotted in Fig. 2(a) for $L= \pi$ and in Fig. 2(b) for $L = 9 \pi$, for $W_0 = 4$. Making use of the analytic continuation of Mathieu function given in eq. (\ref{mathieu-cont}) above, we plot a typical scattering state in Fig. 2(c), for $E=5$. For this parameter region denoting unbroken ${\cal{PT}}$ symmetry, when the ${\cal{PT}}$ symmetric optical potential can be mapped to a Hermitian potential by a similarity transformation \cite{bikash-PLA}, the scattering is observed to be normal : $|R_{L,R}|^2 \leq 1, \ |T|^2 \leq 1$. (Subscripts $R$ and $L$ denote incidence from right and left, respectively.) As is common in non Hermitian quantum mechanics, the scattering coefficients do not add up to unity : $|R_{L,R}|^2 + |T|^2 \neq 1$. On the contrary, they satisfy the generalized unitarity relation discussed in \cite{ge-pra} :
\begin{equation}\label{conserve}
    \displaystyle \mid \mathbb{T} - 1 \mid = \sqrt{\mathbb{R_R} \mathbb{R_L}} \ , \qquad \mathbb{T} = |T|^2  , \ \mathbb{R_{R,L}} = |R_{R,L}|^2
\end{equation}
which reduces to
\begin{equation}\label{unitary-pm}
 \begin{array}{lll}
    |T| ^2 &+& \sqrt{ |R_R|^2 |R_L| ^2} = 1 \ , \qquad |T|^2 \leq 1 \\ \\
    |T| ^2 &-& \sqrt{ |R_R|^2 |R_L| ^2} = 1 \ , \qquad |T|^2 > 1 \\
    \end{array}
\end{equation}
Since transmittance is normal in this parameter regime, the first of eq. (\ref{unitary-pm}) is obeyed here, as is evident in Figures 2(a) and 2(b). With increase in the size of the periodic structure $L$, the only difference observed is that the number of oscillations of $|R_{R,L}|^2$ and $|T|^2$ increases, as shown in Fig. 2(b), for $L=9 \pi$.

\vs{.3cm}

\begin{figure}[ht]
\begin{center}
\includegraphics[width=4.5 cm,height=3.4 cm]{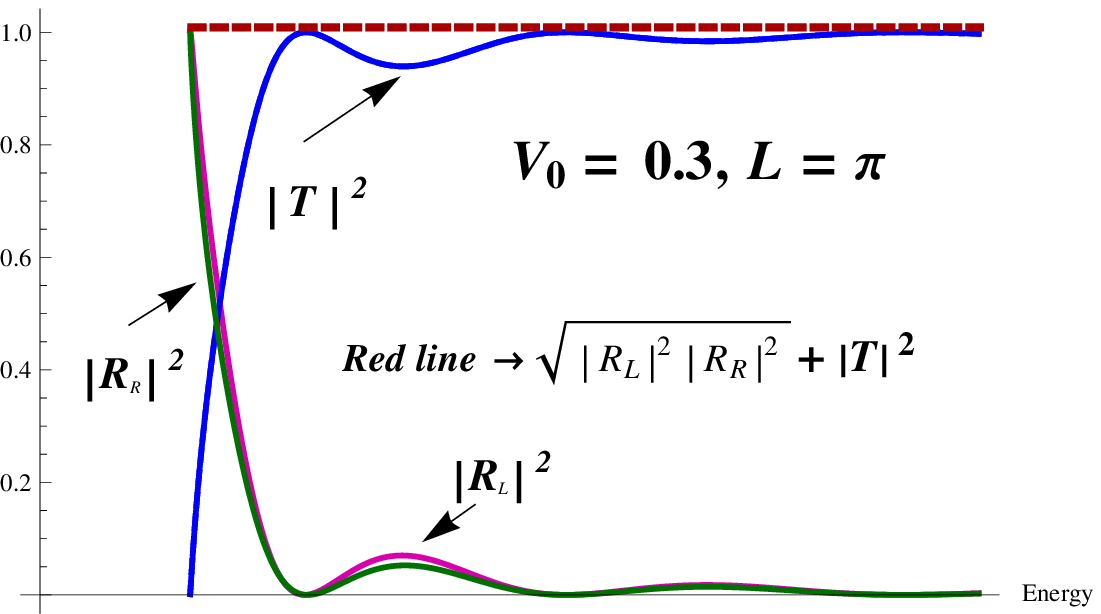} ~~~~~ \includegraphics[width=4.5 cm,height=3.4 cm]{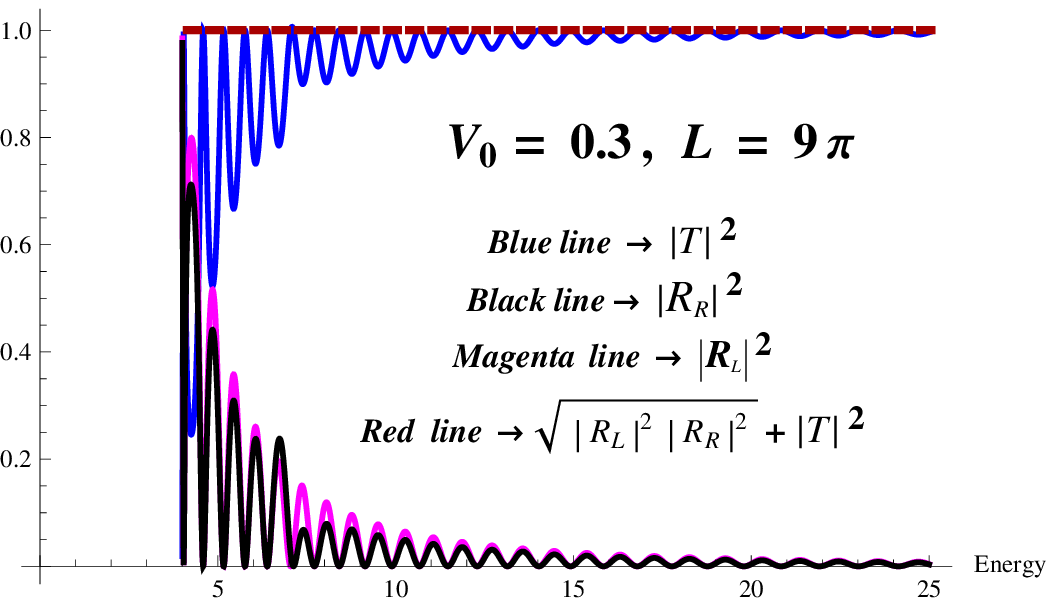} ~~~~~
\includegraphics[width=4.5 cm,height=3.4 cm]{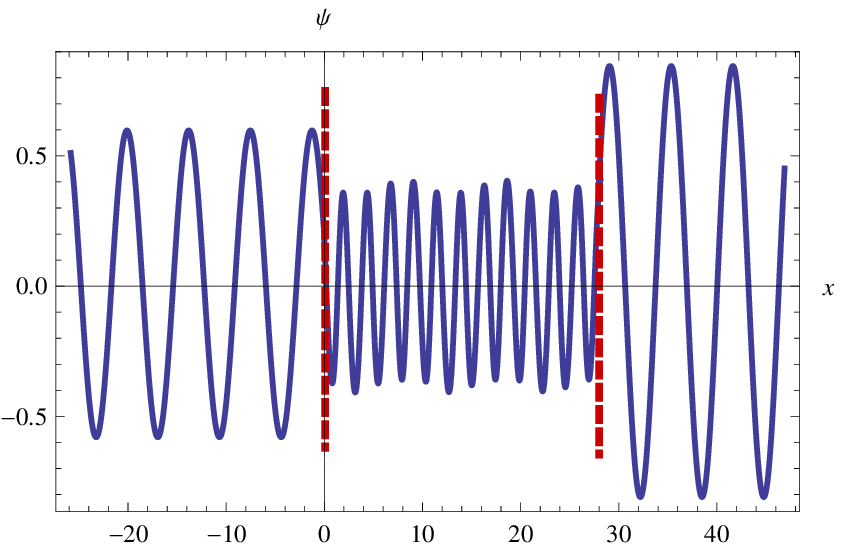} \\
\caption{{Color online : Plot showing $|T|^2$, $|R_R|^2$ and $|R_L|^2$ w.r.t. Energy for $V_0 = 0.3$, $W_0 = 4$} ~~ (a) {\small $L=\pi$ }
    ~~ (b) {\small $L=9 \pi$ } ~~ (c) Plot showing $ Re \ \psi$ for $E = 5$, $L=9 \pi$; red dashed lines show the bounding walls at $L=0$ and $L=9 \pi$}
\end{center}
\end{figure}

\vspace{.5cm}

\noindent {\bf Case 2 : $V_0 > 0.5$}

\vs{.2cm}

Changing the variable to
\begin{equation}\label{y-more}
    \hat{y} = x - \displaystyle \frac{i}{2} \ \tanh  ^{-1} \left( \frac{1}{2 V_0} \right)
\end{equation}
along with the transformation
\begin{equation}\label{yb}
    \bar{y} = \displaystyle \frac{\pi}{4} - \hat{y}
\end{equation}
eq. (\ref{schro-eq}) again reduces to the Mathieu equation
\begin{equation}\label{math-sin}
    \displaystyle \frac{d^2 \psi}{d\bar{y}^2} + \left[ \bar{a} - 2\bar{q} \ \cos 2\bar{y} \right] \psi = 0
\end{equation}
with characteristic value
\begin{equation}\label{a-sin}
    \displaystyle \bar{a} =  \frac{W_0}{2} - \beta
\end{equation}
and
\begin{equation}\label{qb}
    \displaystyle \bar{q} =  i \frac{W_0}{4} \sqrt{4 V_0 ^2 - 1}
\end{equation}
Thus, within the complex periodic structure, the solutions to eq. (\ref{math-sin}) are again given by the Mathieu functions
\begin{equation}\label{psi-cos}
    \displaystyle \psi _{in} (\bar{y}) = B_1 \ Me _{\nu} ^1 \left( \bar{a}, \bar{q}, \bar{y} \right) \ +
    \ B_2 \ Me _{\nu} ^2 \left( \bar{a},\bar{q},\bar{y} \right)
\end{equation}
The analytical expressions for the scattering amplitudes obtained in this case are similar to those given in eq. (\ref{t}) to (\ref{sigma2}) above, with $B_1 = A_1 , \ B_2 = A_2$.
However, in this case $M^1 (y_0) , \  M^1 (y_L) $  denote the values of the Mathieu function $ Me _{\nu} ^1 \left( \bar{a}, \bar{q}, \bar{y} \right)$ at $x=0$ and $x=L$, and $ M^2 (y_0) , \ M^2 (y_L) $ etc. denote the values of the Mathieu function $ Me _{\nu} ^2 \left( \bar{a}, \bar{q}, \bar{y} \right) $ at $x=0$ and $x=L$. $M^{1 \ \prime} (y_0),\ M^{1 \ \prime} (y_L), \ M^{2 \ \prime} (y_0) , \ M^{2 \ \prime} (y_L) $ represent the corresponding derivatives wrt $x$, at $x=0$ or $x=L$ as the case may be.

The reflectance and transmittance for this particular regime of $V_0$ are plotted in Fig. 3, for $V_0 = 0.8, \ W_0 = 4$ for different values of length $L$. This is the region of spontaneously broken ${\cal{PT}}$ symmetry, when energies turn complex and the bands start merging together. The abrupt ${\cal{PT}}$ phase transition at the critical point $V_0 ^{th} = 0.5$ has interesting manifestations in the scattering spectrum, too, as shown in Fig. 3. From normal scattering for $V_0 < 0.5$, scattering turns anomalous for $V_0 > 0.5$ ($|R_L|^2, \ |T|^2 $ not necessarily less than unity) when the particle enters the device from the emissive (left) side ($Im [V(x)] > 0$). However, reflection remains normal ($|R_R|^2 \leq 1$) when the particle enters the device from the absorptive (right) side ($Im [V(x)] < 0$). The observation is the same whether we consider a single period --- Fig. 3(a), or multiple periods --- Fig. 3 (b), (c), (d). Additionally, the scattering coefficients satisfy the generalized unitarity relation (\ref{conserve}) discussed in \cite{ge-pra} : in this case they satisfy the second equation in (\ref{unitary-pm}). In this case also one can plot the wave function in the entire device, making use of the relation (\ref{mathieu-cont}). Since the qualitative picture is similar to that plotted in Fig. 2(c), we are omitting it here.

\vs{.3cm}

\begin{figure}[ht]
\begin{center}
\includegraphics[width=5 cm,height=3.6 cm]{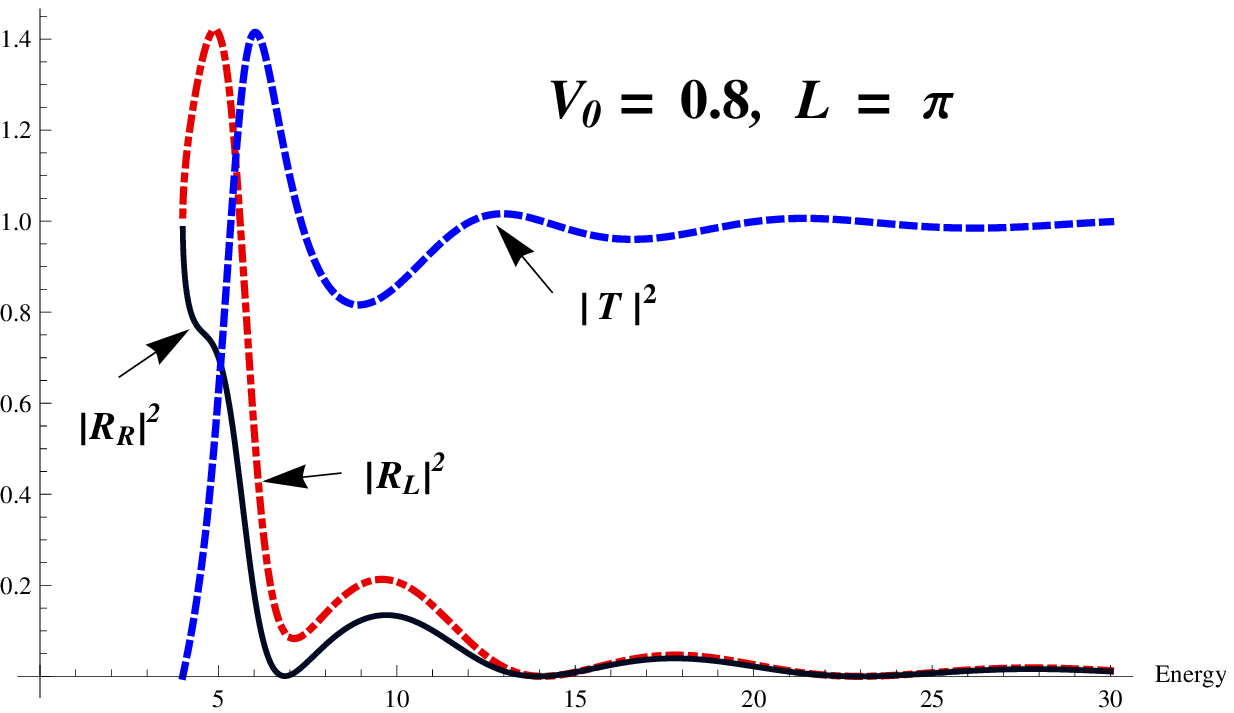} ~~~~~ \includegraphics[width=5 cm,height=3.6 cm]{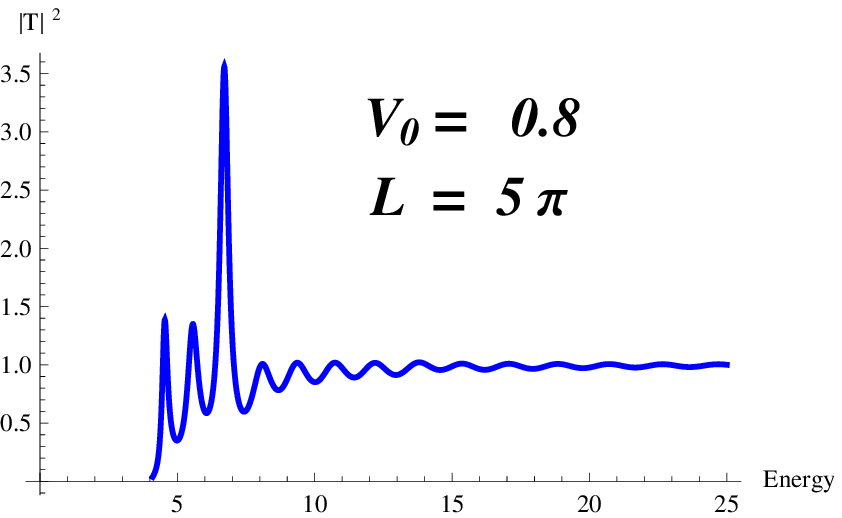} \\ \includegraphics[width=5 cm,height=3.6 cm]{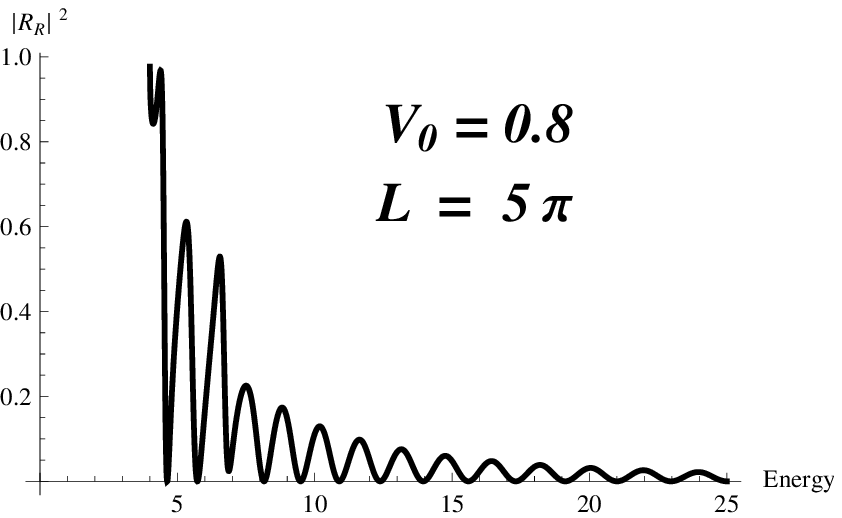} ~~~~~ \includegraphics[width=5 cm,height=3.6 cm]{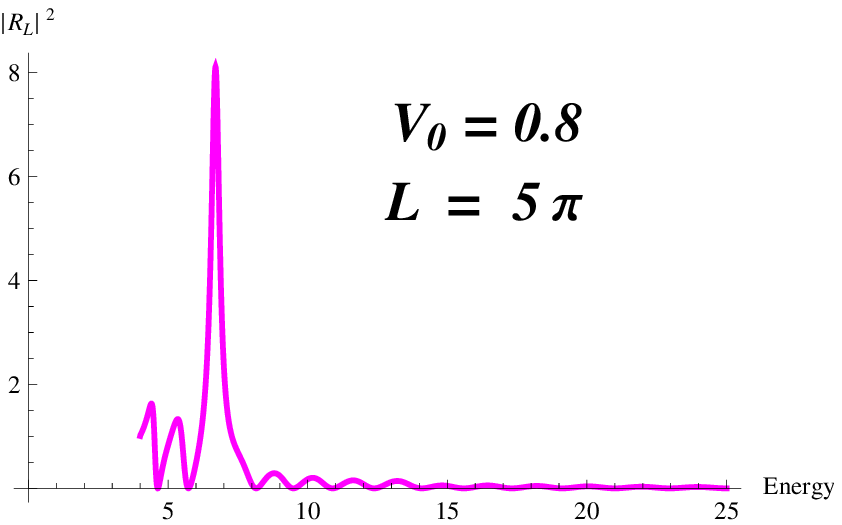} \\
\caption{{Color online : Plot showing $|T|^2$, $|R_R|^2$ and $|R_L|^2$ w.r.t. Energy for $V_0 = 0.8$, for $W_0 = 4$} ~~ (a) {\small $L=\pi$ }
    ~~ (b) {\small $L=5 \pi$ } ~~ (c) {\small $L=5 \pi$ } ~~ (d) {\small $L=5 \pi$ }  }
\end{center}
\end{figure}

\vspace{.3cm}

Another revealing feature in this parameter regime is the observation of spectral singularity (SS). Also known as zero-width resonance, spectral singularities are typically associated with the blowing up of $|T|^2$ and $|R_{R,L}|^2$  \cite{mostafazadeh-ss}. Figures 4(a), 4(b) and 4(c) give the 3-d plot for divergent $|T|^2$ at SS, for $L=\pi, \ 2 \pi $ and $ 5 \pi $ respectively. $|R_{R,L}|^2$ show similar behaviour. For the confined potential studied here, for $V_0 > 0.5$, the number of SS is observed to be infinite. With increasing $L$, the oscillations in $|T|^2$ and $|R|^2$ become more prominent. To the best of our knowledge this analytical result is totally new, not reported till date. Figures 5(a), 5(b) and 5(c) give some typical values of energy $E_{SS}$ where SS takes place --- at $E_{SS}= 5.61$ for $V_0 = 1.115, \ W_0 = 4, L = \pi$, at  $E _{SS} = 39.72$ for $V_0 = 5.794, \ W_0 = 4, \ L = 2 \pi$,  at  $E _{SS}= 9.545$ for $V_0 = 2.70, \ W_0 = 4, \ L = 5 \pi$.

It is worth mentioning here that wave scattering from complex potential barriers enables a finite number of spectral singularities in the continuous spectrum  \cite{samsonov-JPA-L,mostafa-PRL}. However, the potential considered in this work was studied in ref. \cite{longhi-ss} for the unconfined case, for the particular value $V_0 = 0.5$. Identifying SS with secular growth of the wave amplitude, the author observed that the number of SS was countable but infinite. In our present analytical analysis for the confined potential, we consider all possible parameter values of $V_0$. Associating SS with divergent scattering amplitudes, we, too, observe infinite number of spectral singularities.


\pagebreak

\begin{figure}[hp]
\begin{center}
\includegraphics[width=4.5 cm,height=4 cm]{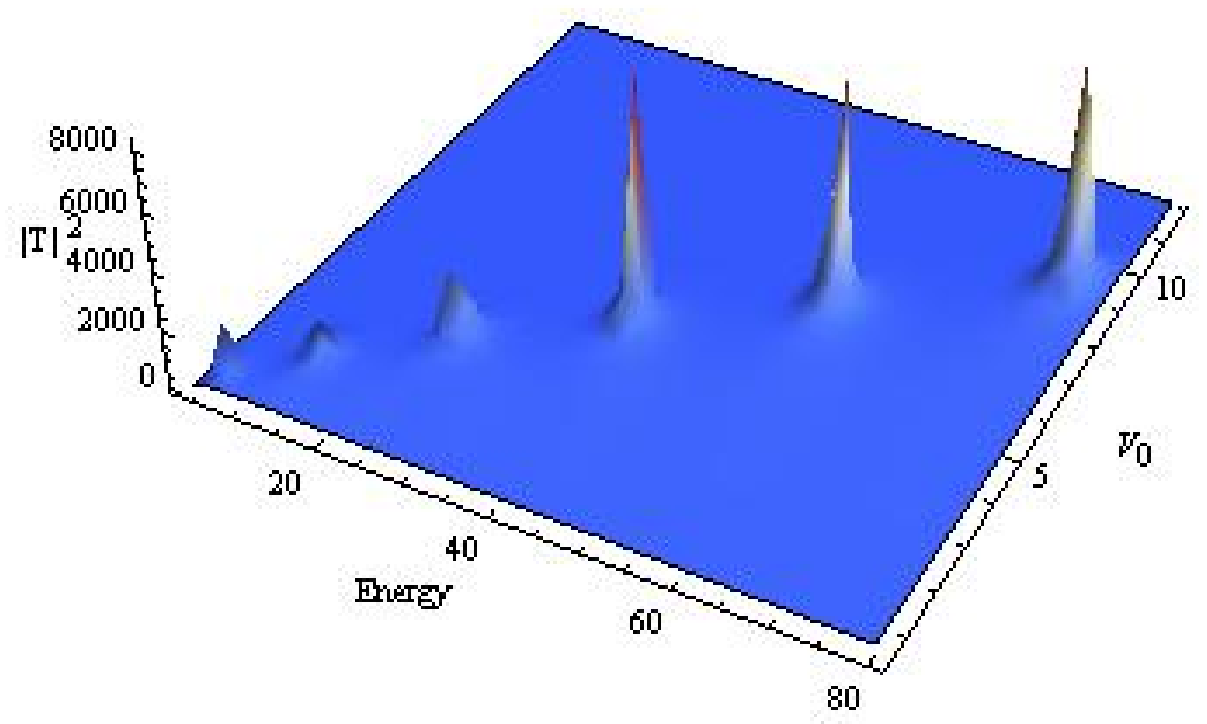} ~~~~~ \includegraphics[width=4.5 cm,height=4 cm]{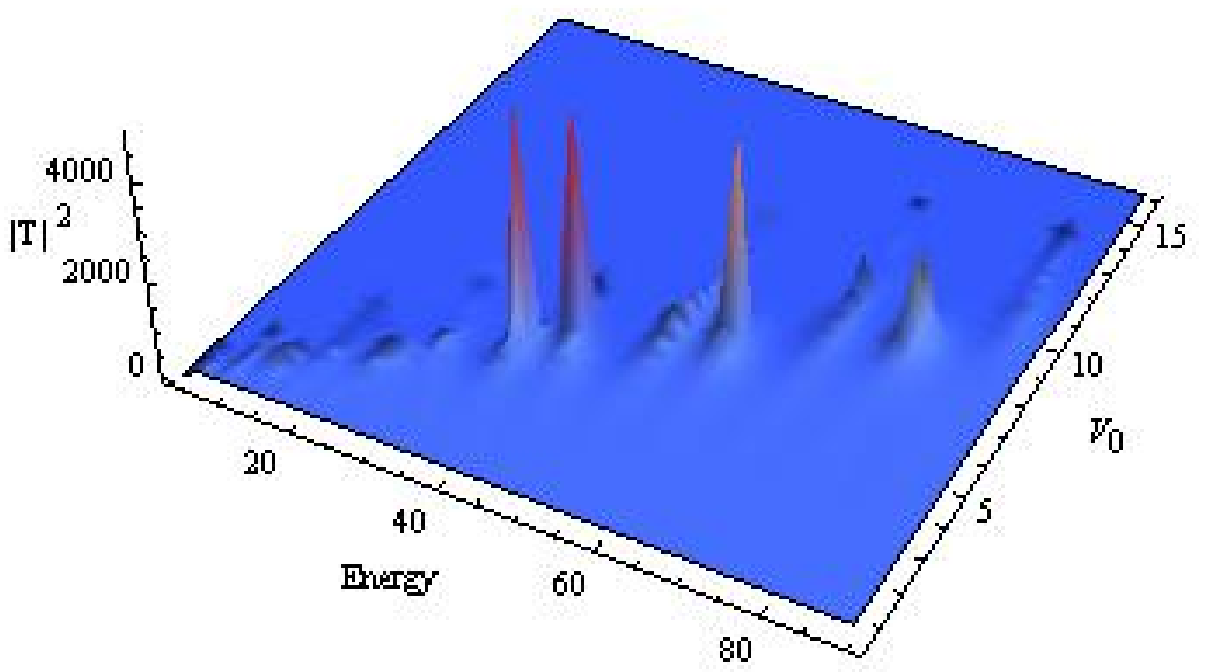} ~~~~~ \includegraphics[width=4.5 cm,height=4 cm]{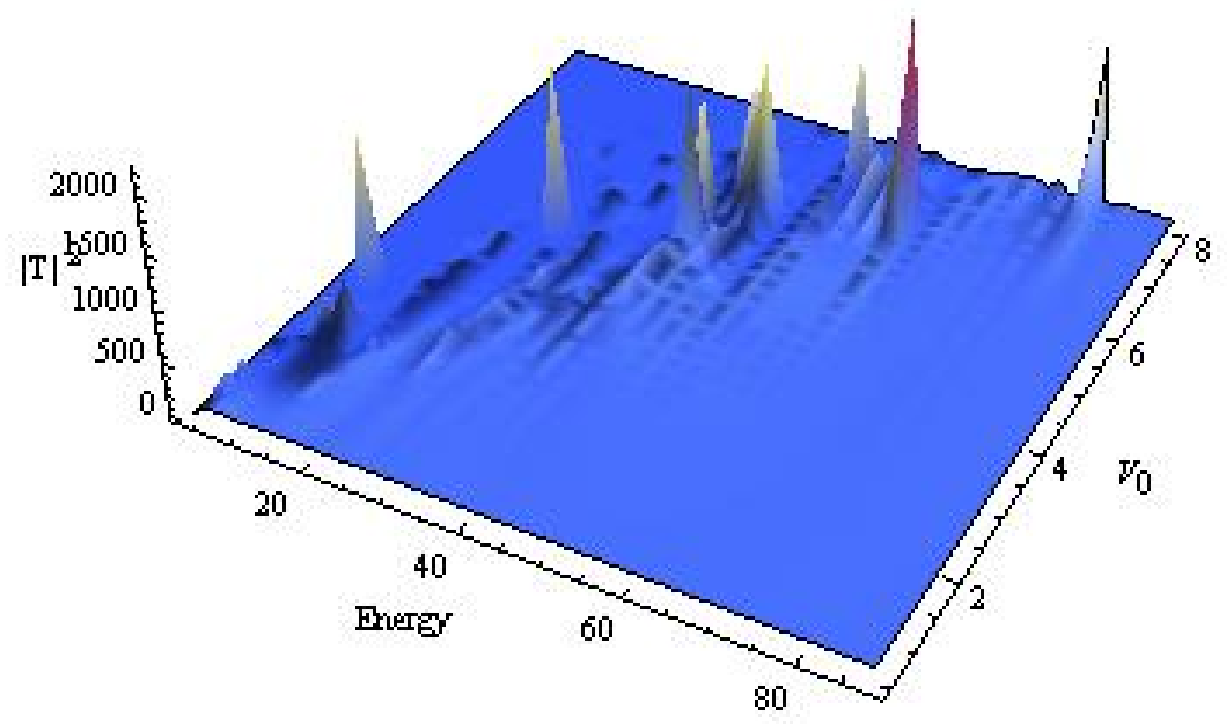} \\
\caption{{Color online :  3D Plot showing multiple spectral singularity at various points of $V_0$, for $W_0 = 4$. Plots are for $|T|^2$; $|R_{R,L}|^2$ show similar behaviour} ~~~~~ (a) { \small $L= \pi$} ~~~~~ (b) {\small $L=2 \pi$ }  ~~~~~ (c) {\small $L=5 \pi$ }}
\end{center}
\end{figure}

\vspace{.3cm}

\begin{figure}[h]
\begin{center}
\includegraphics[width=4 cm,height=3 cm]{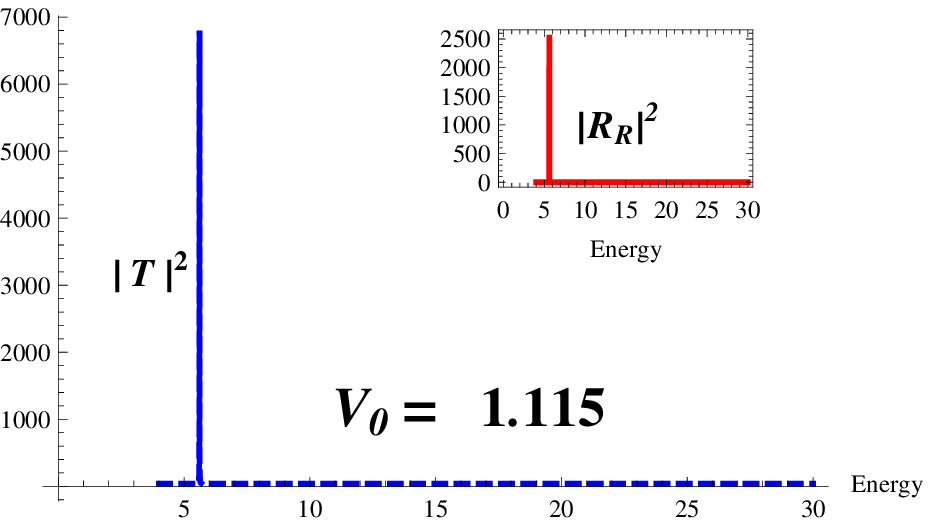} ~~~~~ \includegraphics[width=4 cm,height=3 cm]{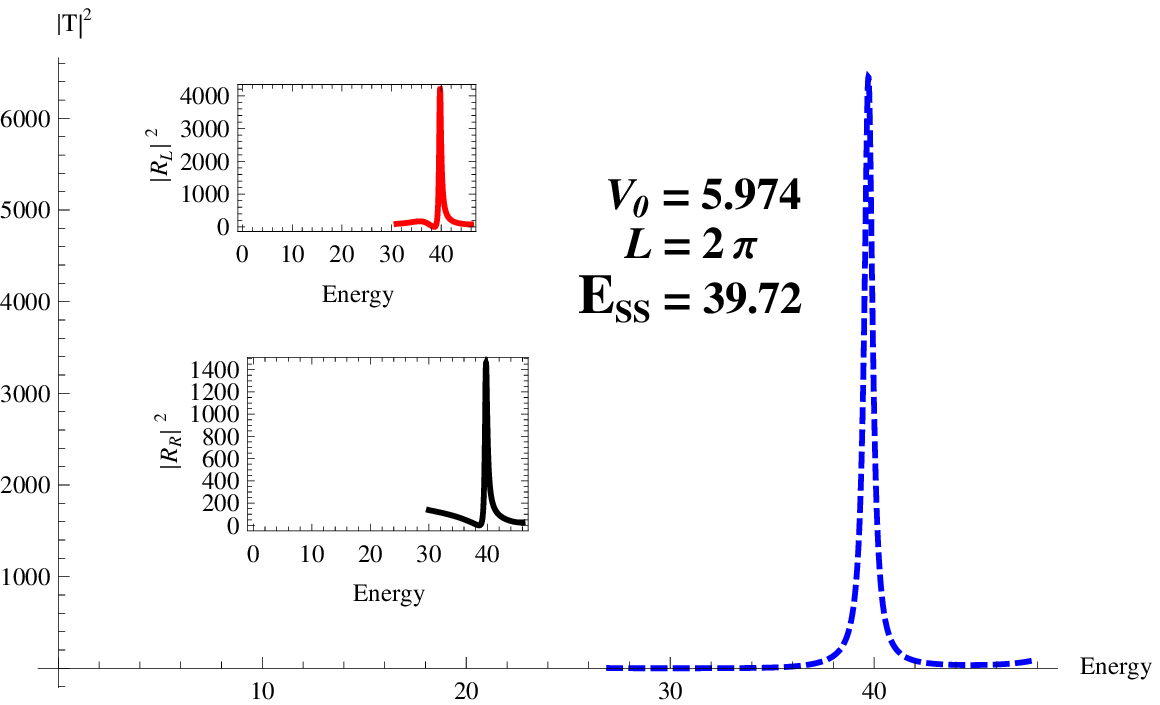}  ~~~~~ \includegraphics[width=4 cm,height=3 cm]{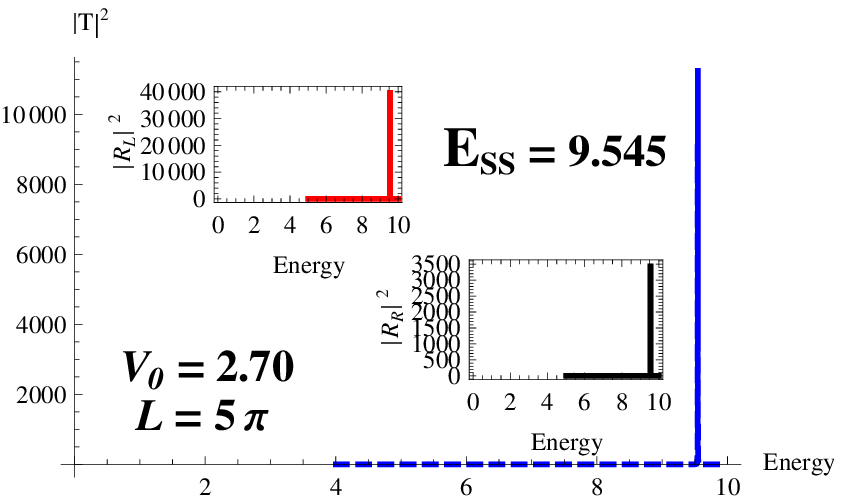} \\
\caption{{Color online : Plot showing blowing up of $|T|^2$ at spectral singularity, inset plots show the same for $|R|^2$ } ~~~~~ (a) {\small SS at $E_{SS}=5.61$, $V_0 = 1.115$ for $W_0 = 4, \ L = \pi$} ~~~~~ (b) {\small SS at $E_{SS}=39.72$, $V_0 = 5.974$ for $W_0 = 4, \ L = 2 \pi$ }  ~~~~~ (c) {\small SS at $E_{SS}=9.545$, $V_0 = 2.70$ for $W_0 = 4, \ L = 5 \pi$ }}
\end{center}
\end{figure}

\vspace{.5cm}


\noindent {\bf Case 3 : $V_0 = 0.5$}

\vs{.2cm}

Transforming $x$ to variable
\begin{equation}\label{y-more}
    \xi = \displaystyle i \sqrt{\frac{W_0}{2}} e^ {ix}
\end{equation}
eq. (\ref{schro-eq}) reduces to the Bessel differential equation
\begin{equation}\label{bessel}
    \displaystyle \xi ^2 \ \frac{d^2 \psi}{d\bar{\xi}^2} + \xi \ \frac{d \psi}{d \xi}
    + \left( \xi ^2 - \kappa ^2 \right) \psi = 0
\end{equation}
where
\begin{equation}\label{kappa}
    \displaystyle \kappa =  \sqrt{ \frac{W_0}{2} - \beta}
\end{equation}
A case similar to this was studied in \cite{jones-jpa}, where bounding walls were considered at $ \pm L/2 $. However, in that  study the variation of the complex refractive index was taken in the transverse direction, whereas we consider the variation in the longitudinal direction.

\vspace{.5cm}

\noindent Now, the solutions to eq. (\ref{bessel}) in the region $0 \leq x \leq L$ are given by the Bessel functions
\begin{equation}\label{bessel-fn}
    \displaystyle \psi _{in} (\xi) = C_1 \ J_{\kappa} (\xi) \ + \ C_2 \ J_{- \kappa}  (\xi)
\end{equation}
Mathematically, one need not consider the case $V_0=0.5$  separately. One can easily check that
    $$ {\rm{lim.}} \ V_0 \rightarrow 0.5-0  , \  \displaystyle \frac{W_0}{2} \left[ 1 + U_1 (x) \right]
     \rightarrow  \displaystyle \frac{W_0}{2} \left( 1 +  e^{2 i x} \right) $$
$$ {\rm{lim.}} \ V_0  \rightarrow  0.5+0  , \ \displaystyle \frac{W_0}{2} \left[ 1 + i U_2 (x) \right]
        \rightarrow  \displaystyle \frac{W_0}{2} \left( 1 +  e^{2 i x} \right) $$
Thus the first and third expressions of (\ref{v}) both tend to the second expression given in (\ref{v}). \\
As $x$ goes from $0$ to $L$, continuation onto subsequent sheets is achieved by using the formula \cite{handbook}
\begin{equation}\label{bessel-cont}
    \displaystyle J_{\kappa} \left( y e^{i n \pi} \right) = \displaystyle e^{i n \kappa \pi} J_{\kappa} (y)
\end{equation}
This expression is used to plot the scattering solutions in the entire device, shown later in Fig. 6(c) for $E=5.6$, $L=4 \pi$.


\vspace{.5cm}

\begin{figure}[h]
\begin{center}
\includegraphics[width=4 cm,height=3 cm]{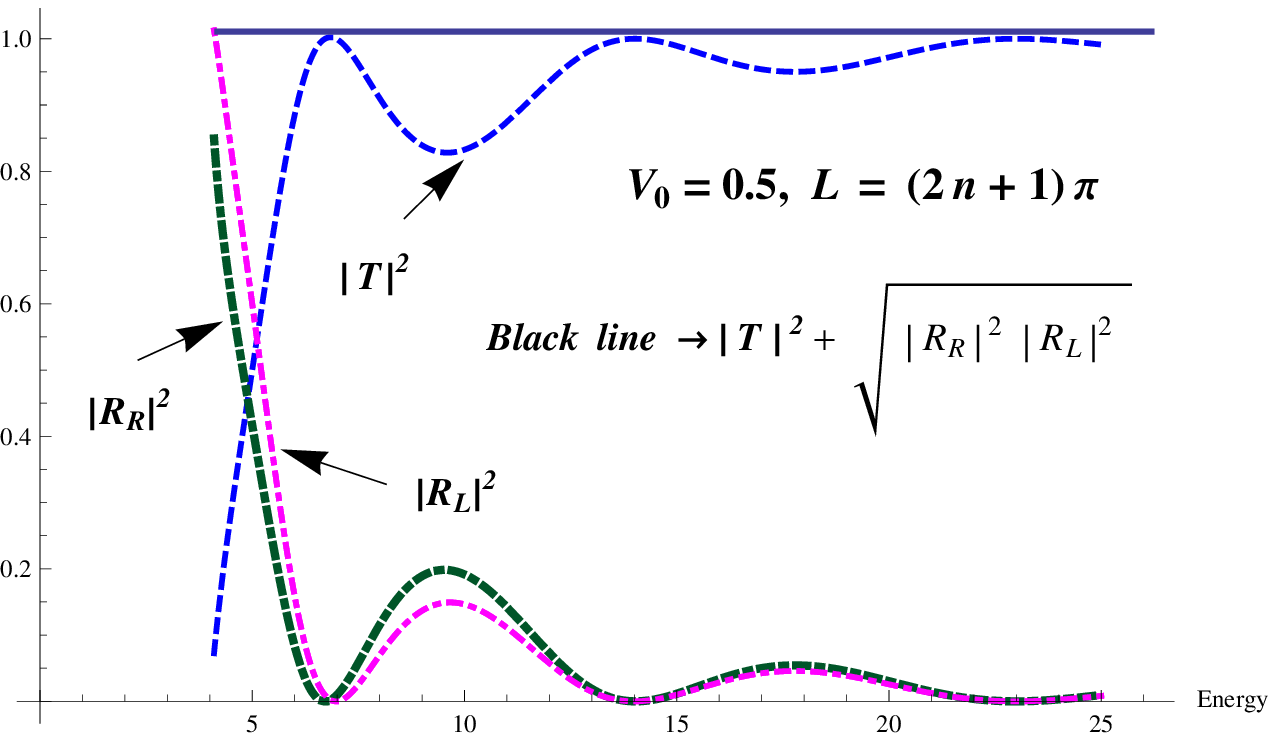} ~~~~~~ \includegraphics[width=4 cm,height=3 cm]{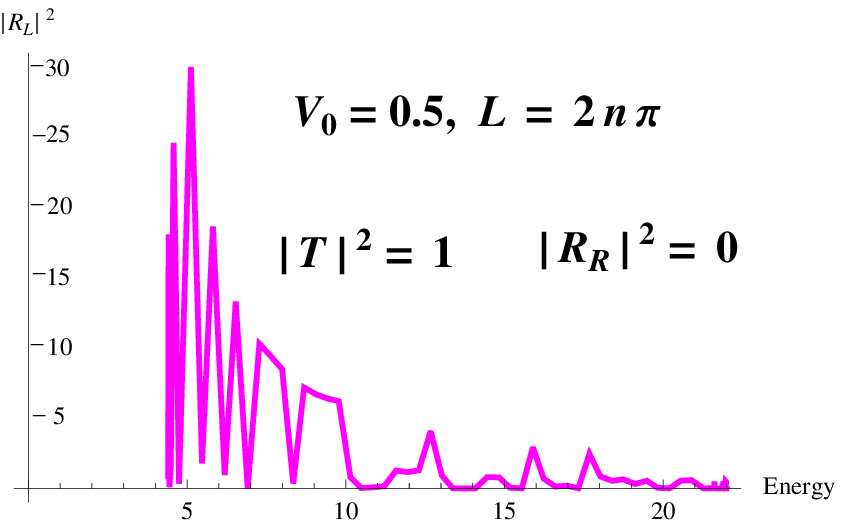} ~~~~~~ \includegraphics[width=4 cm,height=3 cm]{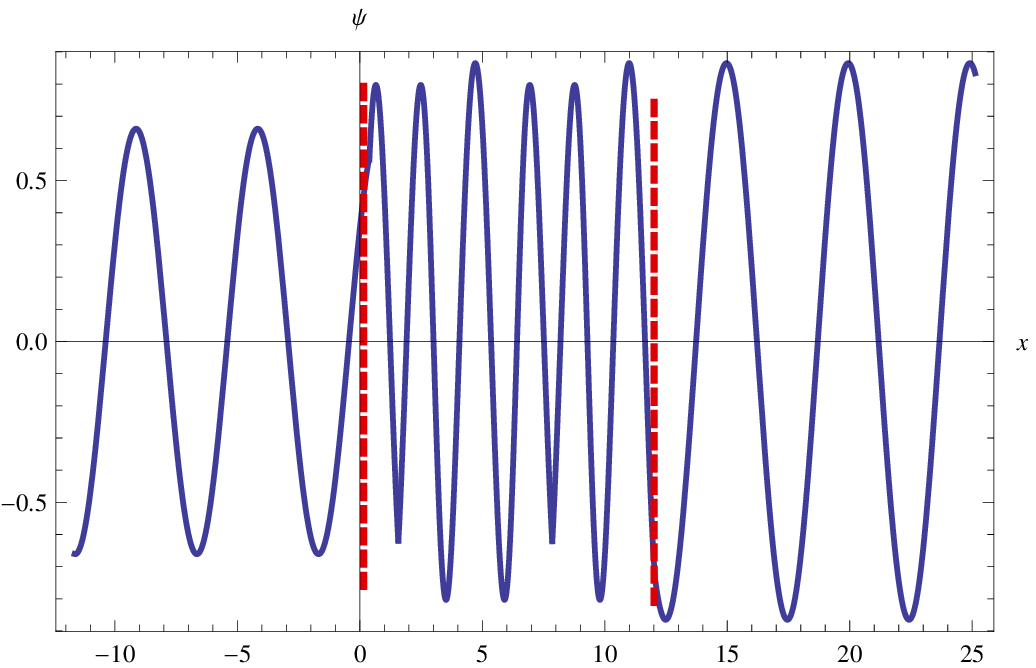} \\
\caption{{Color online : Plot showing $|T|^2$ and $|R_{L,R}|^2$ w.r.t. Energy for $V_0 = 0.5$, for $W_0 = 4$} ~~ (a) {\small $L = (2n+1) \pi$ }
    ~~ (b) {\small $L = 2 n \pi$ } ~~ (c) Plot showing $Re \ \psi$ for $E=5.6$, $L=4 \pi$; red dashed lines show the bounding walls at $L=0$ and $L=4 \pi$}
\end{center}
\end{figure}

\vspace{.5cm}

\noindent Also when $Im \ [|z|]  \rightarrow  \infty$, Mathieu function becomes Hankel function.
Using the notation used by Dunster \cite{dunster}
\begin{equation}\label{asymp}
    \displaystyle M_{\nu} ^{(3,4)} (z,q) \  \sim  \ H_{\nu} ^{(1,2)} \left( 2 \sqrt{q} \xi \right)
\end{equation}
In the present case $ \xi = - \ \cos (y) $, where $y$ is given by eq. (\ref{y-less}), $q$ by eq. (\ref{q-cos}), and
\begin{equation}\label{hankel}
    H_{\nu} ^{(1,2)} = J_{\nu} (z) \pm i Y_{\nu} (z)
\end{equation}
Noting that
\begin{equation}\label{bessel-y}
    Y_{\nu} (z) = \displaystyle \frac{J{\nu} (z) \ \cos (\nu \pi) - J_{- \nu} (z) }{\sin (\nu \pi)}
\end{equation}
one gets back the solution (\ref{bessel-fn}) when $V_0  \rightarrow  0.5$. Similar is the situation when $V_0  \rightarrow 0.5 + 0$.

\vspace{.5cm}

\noindent The analytical expressions for the scattering amplitudes are of the form
\begin{equation}\label{t-b}
    T = \displaystyle \frac{e^{-ikL}}{2ik} \ \left[ C_1 \left( - \ \sqrt{W_0/2} \ e^{iL} J_{\kappa} ^{\prime} (\xi _L) + i k J_{\kappa} (\xi _L) \right) + \displaystyle  C_2 \left( ik J_{- \kappa} (\xi _L)   - \ \sqrt{W_0/2} \ e^{iL} J_{- \kappa} ^{\prime} (\xi _L) \right) \right]
\end{equation}
\begin{equation}\label{rR-b}
    R_R =  \displaystyle \frac{1}{2ik} \ \left[ C_1 \left( ik J_{\kappa} (\xi _0) + \ \sqrt{W_0/2} \ J_{\kappa} ^{\prime} (\xi _0) \right) + \displaystyle C_2 \left( ik J_{- \kappa} (\xi _0) +  \ \sqrt{W_0/2} \ J_{- \kappa} ^{\prime} (\xi _0) \right) \right]
\end{equation}
\begin{equation}\label{rL-b}
    R_L =  \displaystyle \frac{1}{2ik \ e^{ikL} } \ \left[ C_{11} \left( ikJ_{\kappa} (\xi _L) - \ \sqrt{W_0/2} \ e^{iL} J_{\kappa} ^{\prime} (\xi _L) \right) + \displaystyle  C_{22} \left( ik J_{- \kappa} (\xi _L)  - \ \sqrt{W_0/2} \ e^{iL} J_{- \kappa} ^{\prime} (\xi _L) \right)  \right]
\end{equation}
\begin{equation}\label{c1}
    C_1 = \displaystyle \frac{2ik}{\sigma _3}  \ , \ C_2 = - C_1 C_{11}
\end{equation}
\begin{equation}\label{sigma3}
    \sigma _3 = \displaystyle  - \sqrt{W_0/2} \ J_{\kappa} ^{\prime} (\xi _0) \ + \ ik J_{\kappa} (\xi _0) -
     C_{11} \left(  -  \sqrt{W_0/2} \ J_{- \kappa} ^{\prime} (\xi _0) +ik J_{- \kappa} (\xi _0) \right)
\end{equation}
\begin{equation}\label{c11}
    C_{11} = \displaystyle \frac{ \ \sqrt{W_0/2} \ e^{iL} J_{\kappa} ^{\prime} (\xi _L) + ik J_{\kappa} (\xi _L) }{ \ \sqrt{W_0/2} \ J_{- \kappa}^{\prime} (\xi _L)  + ik J_{- \kappa} (\xi _L)}
\end{equation}
\begin{equation}\label{c22}
    C_{22} = \displaystyle \frac{ \ \sqrt{W_0/2} \ e^{iL} J_{\kappa}^{\prime} (\xi _0)  - ik J_{\kappa} (\xi _0) }{ \ \sqrt{W_0/2} \ J_{- \kappa} ^{\prime}(\xi _0)  -ik J_{- \kappa} (\xi _L)}
\end{equation}
where prime denotes the derivative of the Bessel functions w.r.t. $\xi$.

The plot of the transmission amplitudes at the critical point $V_0 ^{th} = 0.5$ displays very interesting phenomenon. For odd number of periodic cells $L = (2 n + 1 ) \pi$, for low values of energy, the reflection and transmission coefficients oscillate. However, for large energies, transmittance reaches unity and reflectance goes to zero as expected ($|T|^2 \rightarrow 1, \ |R_{L,R}|^2 \rightarrow 0$). This is shown in Fig. 6(a). Furthermore,
if $\kappa = m \pi$, where $m$ is any integer, then using the properties of Bessel functions one can show that $| R_R R_L | = 0$ and $|T|=1$. This feature is also evident in Fig. 6(a). Additionally, the scattering coefficients satisfy the generalized unitarity relation (\ref{conserve}) discussed in \cite{ge-pra} for 1-dimensional ${\cal{PT}}$ symmetric photonic heterostructures : $ |T| ^2 + \sqrt{ |R_R|^2 |R_L| ^2} = 1 $ for normal transmittance $|T|^2 \leq 1$ and $ |T| ^2 - \sqrt{ |R_R|^2 |R_L| ^2} = 1 $ for anomalous transmittance $|T|^2 > 1$. Earlier studies on the band structure of this complex potential have shown that all eigenvalues for every band and every Bloch wave number are real and all the forbidden gaps are open, for $V_0 < 0.5$, while the band gaps disappear at the threshold $V_0 ^{th} = 0.5$ \cite{pt-opt1,pt-opt2,longhi-ss}.

For even number of periodic cells $L=2 n \pi$, we observe an interesting phenomenon --- viz., unidirectional invisibility, similar to the numerical simulations reported in earlier works \cite{pt-opt4}. It is easy to see that at $L=2 n \pi$, $ J _{\pm \kappa} ( \xi_0) = J_{\pm \kappa} ( \xi_L) = J_{\pm \kappa} \ {\rm{(say)}} $, so that eq. (\ref{t-b}) and (\ref{rR-b}) reduce to  $T = \displaystyle e^{-2ikn \pi} \Rightarrow |T|^2 = 1 $  and $R_R = 0$, respectively, whereas the expression for $R_L$ in eq. (\ref{rL-b}) reduces to
\begin{equation}\label{rr-2npi}
  R_L = \displaystyle \frac{1}{2ik} \ \left( ik \ J_{\kappa} - \alpha \ J_{\kappa} ^{\prime} \right) \ \left[ \frac{\alpha \ J_{\kappa} ^{\prime} - i k \ J_{\kappa} }{\alpha \ J_{- \kappa} ^{\prime} - i k \ J_{- \kappa}} +1 \right]
\end{equation}
where $\alpha = \sqrt{W_0/2}$. Thus, for right incident waves the potential appears reflectionless : $|R_R| = 0$, whereas for left incident waves there is finite reflection $|R_L| \neq 0$ as shown in Fig. 6(b), with identically unit transmission in either case. We checked this numerically also and arrived at a similar result. This typical phenomenon of unidirectional invisibility --- zero reflectance from one side and unit transmittance --- is also referred to as an anisotropic transmission resonance (ATR) \cite{ge-pra}. This may be seen as a generalization of the flux-conserving transmission resonances of unitary systems (when $R_L = R_R$). The generalized unitarity relation for 1-dimensional ${\cal{PT}}$ symmetric photonic heterostructures mentioned in eq.(\ref{conserve}) above, follows naturally. We must emphasize that the results presented here are based on exact analytical expressions for $|R_{R,L}|^2 , \ |T|^2$, written in terms of Bessel functions and their derivatives.

Finally, Fig. 7 shows the reflectance and transmittance in the absence of gain-loss modulation --- i.e., the corresponding Hermitian optical potential ($V_0 = 0$). As expected, there is no left-right asymmetry, and the sum of the scattering coefficients always add up to unity. Increasing the size of the periodic structure only increases the number of oscillations in the scattering amplitudes.

\vspace{.5cm}

\begin{figure}[h]
\begin{center}
\includegraphics[width=4.5 cm,height=3 cm]{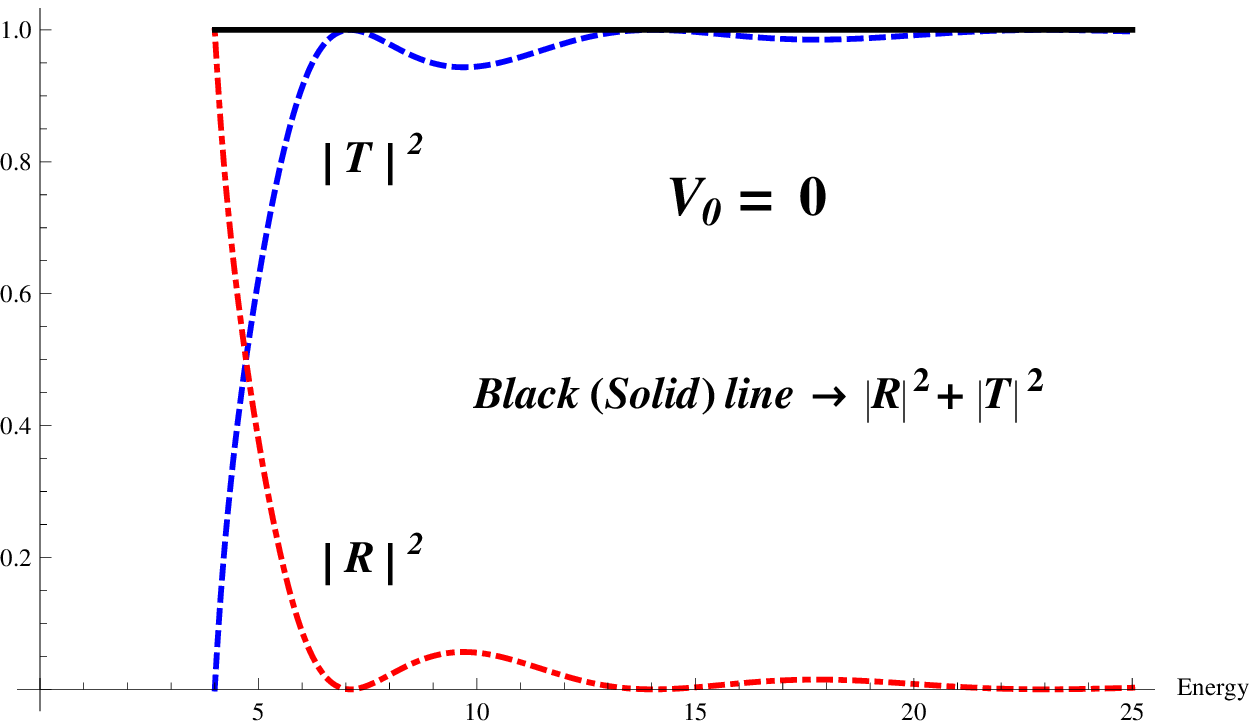} ~~~~~ \includegraphics[width=4.5 cm,height=3 cm]{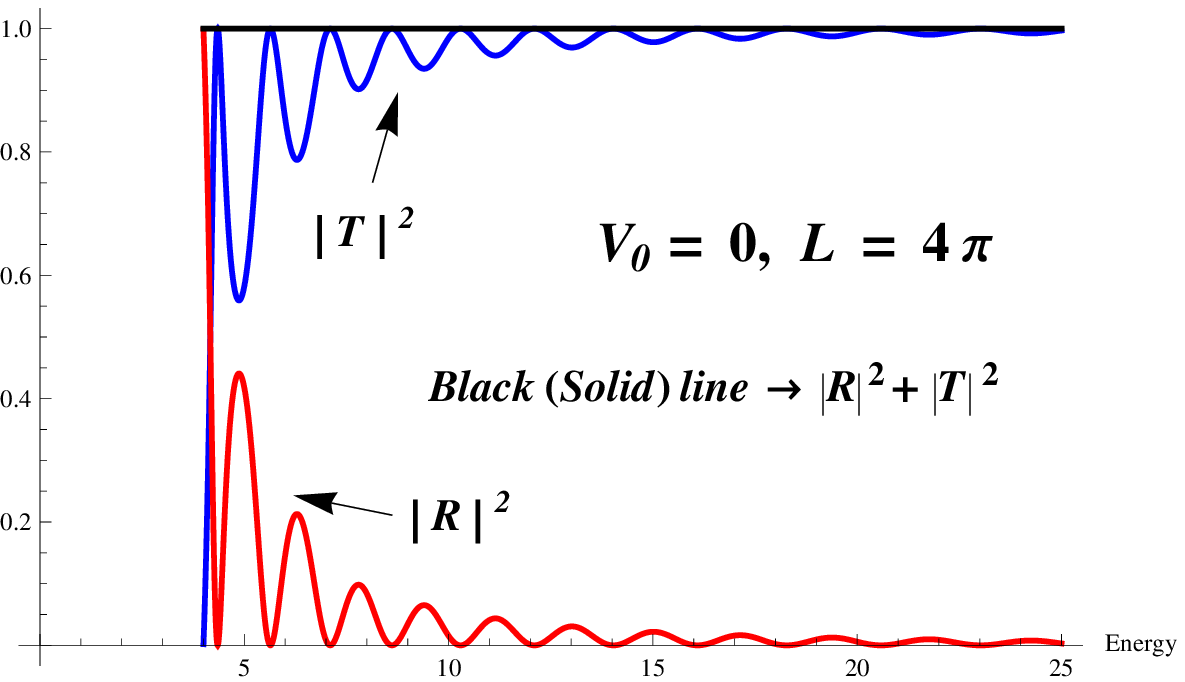} ~~~~~ \includegraphics[width=4.5 cm,height=3 cm]{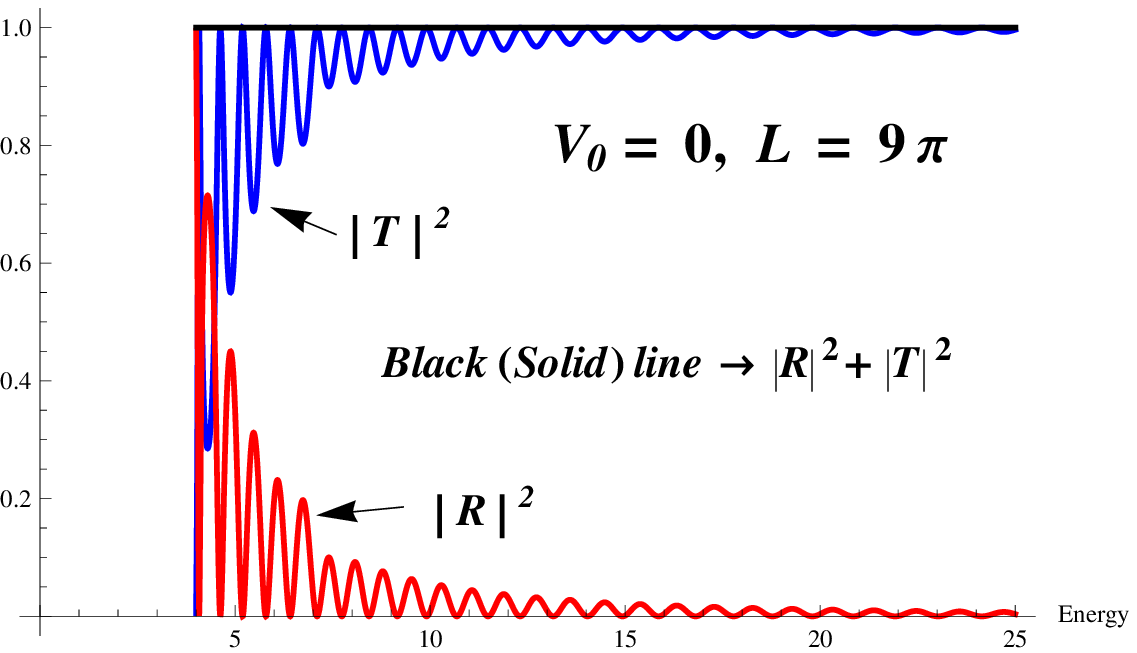}
\caption{{Color online : Plot showing $|R|^2$ and $|T|^2$ at $V_0=0$, for $W_0 = 4$} ~~ (a) { \small $ L = \pi$} ~~
    (b) {\small $L = 4 \pi$} ~~ (c) {\small $L = 9 \pi$}}
\end{center}
\end{figure}

\vspace{.5cm}

\section{Summary}

Motivated by the studies on complex crystals, in this work we investigated a ${\cal{PT}}$ symmetric optical potential of period $\pi$, confined in the region $0 \leq x \leq L$, embedded in a homogeneous medium having uniform potential $W_0$; $V(x) = W_0$ for $x \leq 0$ and $x \geq L$. Our main emphasis was to look for spectral singularities in the continuous spectrum, and unidirectional invisibility, if any. Our probe revealed a number of interesting features. For values of $V_0$ below the critical point $0.5$ the scattering is normal ($|T|^2 , \ |R_{R,L}|^2 \ \leq \ 1$), whereas it turns anomalous beyond this point ($|T|^2 , \ |R_L|^2 $ not necessarily $ \leq \ 1$). Left-right asymmetry, typical of non Hermitian quantum systems is evident in each parameter regime. Whereas $|T_R| = |T_L| = |T|$ (say) in each case, $|R_R| \neq |R_L|$, as observed in other non Hermitian quantum systems. Figures 2 to 7 illustrate our findings. Additionally, this particular model satisfies the modified unitarity relation (\ref{conserve}) developed in \cite{ge-pra}, in each parameter region.

The critical point $V_0 ^{th} = 0.5$ shows interesting behaviour in the scattering spectrum. For odd number
of cells in the periodic structure $L = (2n + 1) \pi$, for low values of energy, the reflection and transmission coefficients
oscillate. For large energies, transmittance reaches unity and reflectance goes to zero : $|T|^2  \rightarrow 1, \ |R_{L,R}| ^2 \rightarrow 0$, for waves incident from either right or left. However, for even number of cells $L = 2n \pi$, the potential appears reflectionless when observed from the absorptive (right) side : $|R_R| =0$. At the same time, one observes enhanced reflection when viewed from the emissive (left) side, transmission being identically unity ($|T|^2 = 1$) in either case. Our exact analytical results are supported by numerical plots as well. This phenomenon, generally referred to as unidirectional invisibility, or anisotropic transmission resonance (ATR), is a generalization of the flux-conserving transmission resonances of unitary systems (when $R_L = R_R$). This was reported in earlier studies as well \cite{pt-opt4,jones-jpa,ge-pra}. However, the results presented here are based on exact analytical expressions.

Another interesting observation in this work is related to spectral singularities, also known as zero-width resonances where $|R_{R,L}|^2$ and $|T|^2$ diverge. These are observed for the parameter regime $V_0 > 0.5$, at particular values of energy $E_{SS}$, and are displayed in Figures 4 and 5. Similar to the observation in ref. \cite{longhi-ss}, we found infinite number of spectral singularities. However, our present study is quite different from that of ref. \cite{longhi-ss} in the sense they studied the unconfined potential, for $V_0 ^{th} = 0.5 $ only, whereas we considered all the parameter regimes $V_0 < 0.5, \ V_0 ^{th} = 0.5, \ V_0 > 0.5$, when the periodic potential is embedded in a homogeneous potential of strength $W_0$, bounded by rigid walls at  $x=0 $ and $x =L$. Furthermore, they described SS as the secular growth of the wave amplitude while in the present study SS are associated with the blowing up of the scattering amplitudes.

Finally, we observe that for this particular structure no accidental flux-conserving points were found : $|R_L| \neq |R_R|$ anywhere.

\section{Acknowledgement}

One of the authors (AS) acknowledges financial assistance from the Department of Science and Technology, Govt. of India, through its grant SR/WOS-A/PS-11/2012. Thanks are also due to H. F. Jones and B. Midya for some helpful comments.


\begin{thebibliography}{999}

\bibitem{pt-expt1} A. Guo, et al., Phys. Rev. Lett. {\bf 103} (2009)
093902.
\bibitem{pt-expt2} C. E. R\"{u}ter, et. al, Nat. Phys. {\bf 6} (2010) 192.
\bibitem{pt-expt3} T. Kottos, Nat. Phys. {\bf 6} (2010) 166.
\bibitem{pt-opt1} Z.H. Musslimani, K.G.Makris, R. El-Ganainy and D.N. Christodoulides, Phys. Rev. Lett. {\bf 100}  (2008) 030402; J. Phys. A
{\bf 41} (2008) 244019.
\bibitem{pt-opt2} K. G. Makris, R. El-Ganainy, D.N. Christodoulides and Z.H. Musslimani, Phys. Rev. Lett. {\bf 100} (2008)
103904 ; Phys. Rev. A  {\bf 81} (2010) 063807; Int. J. Theor. Phys. {\bf 50} (2011) 1019.
\bibitem{pt-opt3} S. Longhi, Phys. Rev. Lett. {\bf 103} (2009) 123601  ; Phys. Rev. A {\bf 82} (2010)
031801(R).
\bibitem{pt-opt4} Z. Lin, et. al, Phys. Rev. Lett {\bf 106} (2011) 213901.
\bibitem{pt-opt5} O. Bendix, R. Fleischmann, T. Kottos and B. Shapiro, Phys. Rev. Lett. {\bf 103} (2009) 030402.
\bibitem{pt-opt6} H. Ramezani, T. Kottos, R. El-Ganainy and D. N. Christodoulides, Phys. Rev. A {\bf 82} (2010) 043803.
\bibitem{pt-opt7} Y. D. Chong, L. Ge, and A. D. Stone, Phys. Rev. Lett. {\bf 106} (2011) 093902.
\bibitem{pt-opt8} H. Schomerus, Phys. Rev. Lett. {\bf 104} (2010) 233601.
\bibitem{longhi-ss} S. Longhi, Phys. Rev. A {\bf 81} (2010) 022102.
\bibitem{mostafazadeh-ss} A. Mostafazadeh, Phys. Rev. A {\bf 80} (2009)
032711.
\bibitem{jones-jpa} H. F. Jones, Jour. Phys. A : Math. Theor. {\bf 45} 135306 (2012).
\bibitem{plyushchay} F. Correa and M. S. Plyushchay, Phys. Rev. D {\bf 86} 085028 (2012).
\bibitem{zafar-2013} Z. Ahmed, Phys. Lett. A {\bf 377} (2013) 957.
\bibitem{mostafa-prl-2013} A. Mostafazadeh, Phys. Rev. Lett. {\bf 110} (2013)
260402.
\bibitem{mostafa-pra-2013} A. Mostafazadeh, Phys. Rev. A {\bf 87} (2013)
063838.
\bibitem{bikash-PLA} B. Midya et. al, Phys. Lett. A {\bf 374} (2010) 2605.
\bibitem{ge-pra} L. Ge, Y. D. Chong and A. D. Stone, Phys. Rev. A {\bf 85} (2012) 023802.
\bibitem{meixner} J. Meixner and F. W. Sch\"{a}fke, {\it Mathieusche Funktionen and Sph\"{a}roidfunktionen}, Springer-Verlag, Berlin, 1954.
\bibitem{handbook} I. Abramowitz and I. A. Stegun, Handbook of Mathematical Functions (1970) (New York: Dover).
\bibitem{dunster} T. M. Dunster, Methods and Applications of Analysis {\bf 1} (1994) 143 (International Press).
\bibitem{samsonov-JPA-L} B. F. Samsonov, J. Phys. A : Math. Gen. {\bf 38} (2005) L397.
\bibitem{mostafa-PRL} A. Mostafazadeh, Phys. Rev. Lett. {\bf 102} (2009) 220402.
\end{thebibliography}
\end{document}